\documentclass[prd,twocolumn,superscriptaddress,altaffilletter,showpacs,nofootinbib]{revtex4}
\usepackage{amssymb}
\usepackage[dvips]{graphicx}
\usepackage{amsmath}

\newcommand{\be}{\begin{equation}}
\newcommand{\ee}{\end{equation}}
\newcommand{\bea}{\begin{eqnarray}}
\newcommand{\eea}{\end{eqnarray}}

\begin{document}

\title{Asymptotic Properties of a Supposedly Regular (Dirac-Born-Infeld) Modification of General Relativity.}

\author{Ricardo Garc\'{\i}a-Salcedo}\email{rigarcias@ipn.mx}\affiliation{Centro de Investigacion en Ciencia Aplicada y Tecnologia Avanzada - Legaria
del IPN, M\'exico D.F., M\'exico.}

\author{Tame Gonzalez}\email{tame@uclv.edu.cu}\affiliation{Departamento de F\'{\i}sica, Universidad Central de Las Villas, 54830 Santa Clara, Cuba.}

\author{Claudia Moreno}\email{claudia.moreno@cucei.udg.mx}\affiliation{Departamento de F\'{\i}sica y Matem\'aticas, Centro Universitario de Ciencias Ex\'actas e Ingenier\'{\i}as,  Av. Revoluci\'on 1500 S.R., Universidad de Guadalajara, 44430 Guadalajara, Jalisco, M\'exico.}

\author{Yunelsy Napoles}\email{yna@uclv.edu.cu}\affiliation{Departamento de F\'{\i}sica, Universidad Central de Las Villas, 54830 Santa Clara, Cuba.}

\author{Yoelsy Leyva}\email{yoelsy@uclv.edu.cu}\affiliation{Departamento de F\'{\i}sica, Universidad Central de Las Villas, 54830 Santa Clara, Cuba.}

\author{Israel Quiros}\email{iquiros@fisica.ugto.mx}\affiliation{Divisi\'on de Ciencias e Ingenier\'ia de la Universidad de Guanajuato, A.P. 150, 37150, Le\'on, Guanajuato, M\'exico.}

\date{\today }

\begin{abstract}
We apply the dynamical systems tools to study the asymptotic properties of a cosmological model based on a non-linear modification of General Relativity in which the standard Einstein-Hilbert action is replaced by one of Dirac-Born-Infeld type. It is shown that the dynamics of this model is extremely rich: there are found equilibrium points in the phase space that can be associated with matter-dominated, matter-curvature scaling, de Sitter, and even phantom-like solutions. Depending on the value of the overall parameters the dynamics in phase space can show multi-attractor structure into the future (multiple future attractors may co-exist). This is a consequence of bifurcations in control parameter space, showing strong dependence of the model's dynamical properties on the free parameters. Contrary to what is expected from non-linear modifications of general relativity of this kind, removal of the initial spacetime singularity is not a generic feature of the corresponding cosmological model. Instead, the starting point of the cosmic dynamics -- the past attractor in the phase space -- is a state of infinitely large value of the Hubble rate squared, usually associated with the big bang singularity.
\end{abstract}

\maketitle

\section{Introduction}

Attempts to modify the Einstein-Hilbert (EH) action of General Relativity (GR) $$S_{EH}=\frac{1}{2\kappa^2}\int d^4x\sqrt{|g|}\left(R-2\Lambda\right),$$ where $R$ is the Ricci curvature scalar, and $\Lambda$- the cosmological constant ($\kappa^2=8\pi m_{Pl}^{-2}=8\pi G$), have been motivated by a number of reasons. In particular, renormalization at one-loop demands that the Einstein-Hilbert action be supplemented by higher order curvature terms \cite{udw}.\footnote{Higher order actions are indeed renormalizable but not unitary \cite{stelle}.} Besides, when quantum corrections or string theory are taken into account, the effective low energy action for pure gravity admits higher order curvature invariants \cite{qstring}. 

There are additional ways to modify the EH GR action. For instance, the one based on the Dirac-Born-Infeld (DBI) procedure for smoothing out singularities \cite{dbiproc,fiorini}.\footnote{The proposal to remove initial as well as final singularities in modified gravity has been given in Ref. \cite{odi}. It was shown there that the addition of a $R^2$ term to otherwise divergent modified gravity makes it regular.} According to this procedure the original Lagrangian density ${\cal L}=\sqrt{|g|}L$ -- whose singularities are to be cured -- is replaced by one of the DBI form: $${\cal L}\;\rightarrow\;{\cal L}_{DBI}=\sqrt{|g|}\mu\left(\sqrt{1+\frac{2L}{\mu}}-1\right),$$ where the scale $\mu$ sets the limit of energy density accessible to the theory.

A combination of the above possible modifications, i. e., a DBI-type action containing an EH term plus additional higher curvature terms within the square root, could supply an additional cosmological scenario where to look for alternative explanations to several phenomena such as inflation and the present speedup of the cosmic expansion.  

Several theories of gravity of this kind have been proposed since long ago in \cite{eddington}, and in more recent years, for instance, in \cite{deser} (see also \cite{comelli}). To be phenomenologically viable, these modifications have to satisfy several physically motivated requirements \cite{deser}:

\begin{enumerate}

\item Reduction to EH action at small curvature,

\item Ghost freedom,

\item Regularization of some singularities (as, for instance, the Coulomb-like Schwarzschild singularity), and

\item Supersymmetrizability.

\end{enumerate}
 
The later requirement is quite stringent and, for most purposes, might be excluded. A DBI-type theory that fulfills the above requirements can be based on the following action \cite{comelli}:

\be S_{DBI}=\frac{1}{\kappa^4}\int d^4x\sqrt{|g|}\left(1-\sqrt{1-\alpha\kappa^2 R+\beta\kappa^4 \mathcal{G}}\right),\label{dbi action}\ee where $\mathcal{G}\equiv R^2-4R_{\mu\nu}R^{\mu\nu}+R_{\mu\nu\sigma\upsilon}R^{{\mu\nu\sigma\upsilon}}$ is the Gauss-Bonnet (GB) invariant. It has been demonstrated in Ref.\cite{comelli} that this action has the EH leading term at small curvature, besides, for an appropriate region in the parameter space it shows indications for the cancellation of the Coulomb-like Schwarzschild singularity, and it is ghost-free.

Given that such a regular theory of gravity could have impact on the cosmology of our universe, in this paper we aim at exploring the asymptotic properties of a Friedmann-Robertson-Walker (FRW) cosmological model based on such a non-linear modification of General Relativity.\footnote{The asymptotic properties of a big class of $f(R,{\cal G})$ theories, with the aim to study future singularities has been discussed in \cite{odi1}.} Our study will rely on the use of the standard tools of the dynamical systems (see, for instance, Ref. \cite{ghale}). The present investigation represents a generalization of the study in Ref.\cite{carloni} to include dependence not only on the Ricci scalar but also on the Gauss-Bonnet term, which is one of the few ways to overcome the presence of ghosts. Notwithstanding, we will not make a comparison of the results we obtain in this paper with the results of \cite{carloni}, since the latter refer to a completely different non-linear modification of general relativity having nothing in common with the DBI modification studied here.   

The paper has been organized as it follows. In the next section the basic mathematics of the model (\ref{dbi action}) are exposed, including the derivation of the relevant FRW cosmological equations. The main section III is devoted to the study of the asymptotic properties of the model. A detailed discussion of the results is given in section IV, while brief conclusions are given in the last section V. Here we use natural units ($\kappa^2=8\pi G=\hbar=c=1$).

\section{Equations of Motion}

We start by writing the action (\ref{dbi action}) in a more compact form \cite{comelli}:

\be S_{DBI}=\frac{1}{2\kappa^2}\int d^4x\sqrt{|g|}\;f(R,\mathcal{G}),\label{dbi f}\ee where

\be f(R,\mathcal{G})=\frac{2}{\kappa^2}\left(1-\sqrt{1-\alpha\kappa^2 R+\beta\kappa^4 \mathcal{G}}\right).\label{F}\ee 

Written in the form (\ref{dbi f}) the action (\ref{dbi action}) has the appearance of an $f(R,{\cal G})$-theory of the kind studied, for instance in Ref.\cite{cognola}, so that we can safely apply known results of these investigations.

The equations of motion that can be derived from (\ref{dbi f}) -- plus the addition of a matter action piece -- by varying in respect to the metric $g_{\mu\nu}$ are the following \cite{cognola}:

\bea &&f_R R_{\mu\nu}-\frac{1}{2}g_{\mu\nu}f-[\nabla_\mu\nabla_\nu-g_{\mu\nu}\Box]f_R-4\Box f_{\mathcal{G}}(R_{\mu\nu} \notag \\
&&\;\;\;\;\;\;\;\;\;\;\;\;-\frac{1}{2}g_{\mu\nu}R)+4(R_{\sigma\nu}\nabla^\sigma\nabla_\mu f_{\mathcal{G}}+R_{\sigma\mu}\nabla^\sigma\nabla_\nu f_{\mathcal{G}} \notag\\
&&\;\;\;\;\;\;\;\;\;\;\;\;\;\;\;\;\;\;-g_{\mu\nu}R_{\sigma\tau}\nabla^\sigma\nabla^\tau f_{\mathcal{G}})+2[2R_{\mu\sigma\nu\tau}\nabla^\sigma\nabla^\tau \notag\\
&&\;\;\;\;\;\;\;\;\;\;\;\;\;\;\;\;\;\;-R\nabla_\mu\nabla_\nu]f_{\mathcal{G}}+2f_{\mathcal{G}}(RR_{\mu\nu}-2R_{\mu\sigma}R^\sigma_{\;\nu} \notag\\
&&\;\;\;\;\;\;\;\;\;\;\;\;\;\;+2R^{\sigma\tau}R_{\mu\sigma\tau\nu}+R_{\mu\sigma\tau\rho}R_\nu^{\;\;\sigma\tau\rho})=\kappa^2 T_{\mu\nu},\label{feq}\eea where $f_R\equiv\partial_R f$, $f_{\mathcal{G}}\equiv\partial_{\mathcal{G}}f$, and $\Box\equiv g^{\sigma\tau}\nabla_\sigma\nabla_\tau$. Worth noticing
that, in case the function $f$ were a function of the curvature scalar only, i. e., in case $f=f(R)$ (formally this case corresponds to the limit $\mathcal{G}\rightarrow 0$), the above field equations reduce to the equations of motion of an $f(R)$-theory \cite{frt,faraonirev}. For purpose of comparison with the equations of EH GR, equations (\ref{feq}) can be recast into the form that resembles Einstein's equations \cite{cognola}:

\be R_{\mu\nu}-\frac{1}{2}g_{\mu\nu}R=\kappa^2_{eff}(T_{\mu\nu}+T^{curv}_{\mu\nu}),\label{einstein feq}\ee where $\kappa^2_{eff}=\kappa^2/(f_R-4\Box f_{\mathcal{G}})$ is the effective gravitational coupling, and we have introduced the following curvature (effective) stress-energy tensor acting as an additional source of the field equations:

\bea &&\kappa^2 T_{\mu \nu }^{curv}=(\nabla _{\mu }\nabla _{\nu }-g_{\mu \nu}\Box )f_{R}+\frac{1}{2}g_{\mu \nu }(f-Rf_{R}) \notag\\
&&\;\;\;\;\;-2(2R_{\mu \sigma \nu \tau }\nabla^{\sigma }\nabla ^{\tau }-R\nabla _{\mu }\nabla _{\nu })f_{\mathcal{G}}-4(R_{\sigma \nu }\nabla ^{\sigma }\nabla _{\mu }f_{\mathcal{G}} \notag\\
&&\;\;\;\;\;\;+R_{\sigma\mu }\nabla ^{\sigma }\nabla _{\nu }f_{\mathcal{G}}-g_{\mu \nu }R_{\sigma\tau }\nabla ^{\sigma }\nabla ^{\tau }f_{\mathcal{G}})-2f_{\mathcal{G}}(RR_{\mu \nu } \notag\\
&&\;\;\;\;\;\;\;\;\;\;-2R_{\mu \sigma }R_{\;\nu }^{\sigma}+2R^{\sigma \tau }R_{\mu \sigma \tau \nu }+R_{\mu \sigma \tau \rho }R_{\nu}^{\;\;\sigma \tau \rho }). \label{Tmncurv}\eea

The trace of equation (\ref{feq}) leads to the following constraint:

\bea &&3\Box f_R+R f_R-2f+2R\Box f_{\mathcal{G}} \notag\\
&&\;\;\;\;\;\;\;\;\;\;\;\;\;\;\;+2[\mathcal{G}-2R_{\sigma\tau}\nabla^\sigma\nabla^\tau] f_{\mathcal{G}}=\kappa^2 T.\label{trace}\eea

For a spatially flat FRW universe metric $$ds^{2}=-dt^{2}+a(t)^{2}(dr^{2}+r^{2}(d\theta +\sin ^{2}\theta d\varphi ^{2}),$$ equations (\ref{feq}) can be written in the form of the following set of cosmological equations:

\bea &&6H^{2}f_{R}=Rf_{R}-f-6H\dot{f_{R}} \notag\\
&&\;\;\;\;\;\;\;\;\;\;\;\;\;\;\;\;\;\;\;\;\;\;\;\;-24H^{3}\dot{f_{\mathcal{G}}}+\mathcal{G}f_{\mathcal{G}}+\rho _{m},\label{friedman}\\
&&\;\;4\dot{H}f_{R}= 2H\dot{f}_R+2\dot{f}_{\mathcal{G}}(12H^3-\mathcal{G}/3H) \notag\\
&&\;\;\;\;\;\;\;\;\;\;\;\;\;\;\;\;\;\;\;\;\;\;\;\;\;\;\;\;-2\ddot{f}_R-8H^2\ddot{f}_{\mathcal{G}}-\rho_m,\label{Raychi}\\
&&\;\;\;\;\;\;\;\dot{\rho}_m=-3H\rho_m, \label{conservation}\eea where the dot accounts for derivative with respect to the cosmic time $t$, and for simplicity we are considering dust as background fluid (it could be, for instance, the dark matter component).

Starting with equation (\ref{F}) it is straightforward to obtain the following relationship between the derivatives of the function $f(R,{\cal G})$:

\be \frac{f_\mathcal{G}}{f_R}\equiv\frac{f_{\mathcal{G}R}}{f_{RR}}\equiv A,\label{relation}\ee where the constant $A\equiv-\kappa^2\beta/\alpha=-\beta/\alpha$ (recall that we use the units system where $\kappa^2=1$). In the present paper, for definiteness, we shall choose only non-negative $\alpha\geq 0$ and $\beta\geq 0$. In consequence, the constant $A$ will be a negative magnitude. 

After (\ref{relation}), and recalling that $${\cal G}=24H^2(\dot H+H^2),\;\;R=6(\dot H+2H^2),$$ we are able to rewrite equations (\ref{friedman},\ref{Raychi}) in a more compact yet simpler form:

\bea && 6H^{2}f_{R}(1+4AH^2)=Rf_{R}(1+4AH^2)\nonumber\\
&&\;\;\;\;\;\;\;\;\;\;\;\;\;\;\;\;-f-6H\dot{f_{R}}(1+4AH^2)+\rho _{m},\label{friedman2}\\
&&4\dot{H}f_{R}+16AH\dot{H}\dot{f}_R=2H\dot{f}_R(1+4AH^2) \nonumber\\
&&\;\;\;\;\;\;\;\;\;\;\;\;\;\;\;\;\;\;\;\;\;\;\;\;\;\;-\rho_m-2\ddot{f}_R(1+4AH^2).\label{Raychi2}\eea Recall that, according to (\ref{einstein feq}), the equation (\ref{friedman2}) can be written in the form of a standard Friedmann equation:

\bea &&3H^2=\kappa^2_{eff}(\rho_m+\rho_{curv}),\;\kappa^2_{eff}=\frac{1}{2f_R(1+4AH^2)},\nonumber\\
&&\kappa^2_{eff}\rho_{curv}=\frac{R}{2}-\frac{f}{2f_R(1+4AH^2)}-3H\frac{\dot f_R}{f_R}.\label{friedmann alternative}\eea Notice that written in this latter form, due to the fact that $A<0$, the Friedmann equation blows up at $H=\pm 1/(2\sqrt{-A})=\pm\sqrt{\alpha/\beta}/2$ (it is in fact the effective gravitational coupling $\kappa^2_{eff}$ who does). However this is not pathological since the above is just a convenient way to write the original equations.

Our goal will be to write the above cosmological equations in the form of an autonomous system of ordinary differential equations, so that we could apply the standard tools of the dynamical systems to find the equilibrium configurations that could be associated with relevant cosmological solutions.

\section{Dynamical Systems Study}

The dynamical systems tools offer a very useful approach to the study of the asymptotic properties of the cosmological models \cite{coley}. In order to be able to apply these tools one has to (unavoidably) follow the steps enumerated below.

\begin{itemize}

\item First: to identify the phase space variables that allow writing the system of cosmological equations in the form of an autonomus system of ordinary differential equations (ODE). There can be several different possible choices, however, not all of them allow for the minimum possible dimensionality of the phase space.

\item Next: with the help of the chosen phase space variables, to build an autonomous system of ODE out of the original system of cosmological equations.

\item Finally (some times a forgotten or under-appreciated step): to identify the phase space spanned by the chosen variables, that is relevant to the cosmological model under study.

\end{itemize}

After this one is ready to apply the standard tools of the (linear) dynamical systems analysis (see, for instance Ref. \cite{dwands}). 

We will split our study into two parts. First, in subsection A, we explore the asymptotic properties of the simplest situation when in (\ref{dbi action}) we set $\mathcal{G}=0$. Although in this case the corresponding DBI theory has a ghost, spin-zero degree of freedom, so it does not admit a stable de Sitter background \cite{qln}, this case is easier to handle and could signal to relevant asymptotic behavior with which to compare the results of the more general ghost-free case ($\mathcal{G}\neq 0$), in the second part of our investigation (subsection B).

\subsection{Einstein-Hilbert DBI Case ($\mathcal{G}=0$)}

\subsubsection{Autonomous System of ODE}

\begin{table*}[ht!]\caption[crit]{Properties of the equilibrium points of the autonomous system (\ref{asode}).}
\begin{tabular}{@{\hspace{4pt}}c@{\hspace{14pt}}c@{\hspace{14pt}}c@{\hspace{14pt}}
c@{\hspace{14pt}}c@{\hspace{14pt}}c@{\hspace{14pt}}c@{\hspace{14pt}}c@{\hspace{14pt}}
c@{\hspace{14pt}}c@{\hspace{14pt}}c@{\hspace{14pt}}c@{\hspace{14pt}}c@{\hspace{14pt}}c}
\hline\hline\\[-0.3cm]
$P_i$ &$x$&$y$&$z$&Existence&$\bar\Omega_m$& $\bar\Omega_{curv}$&$w_{eff}$& $q$ \\[0.1cm]\hline\\[-0.2cm]
%%%%%%%%
$P_1$& $0$&$0$&$1$&Always&$0$&$1$&$1/3$&$1$\\[0.2cm]
$P_2$& $0$&$0$&$-1$&"&$2$ &$-1$&$1/3$&$1$\\[0.2cm]
$P_3$& $-1$&$0$&$-3$&"&$3$&$-2$&$1$&$2$\\[0.2cm]
$P_4$& $2$&$4$&$-1$&"&$0$&$1$&$-1$&$-1$\\[0.2cm]
$P_5$& $2$&$4$&$3$&"&$-4$&$5$&$-1$&$-1$\\[0.2cm]
$P_6$& $2$&$9/2$&$0$&"&$0$&$1$&$-1$&$-1$\\[0.2cm]
\hline \hline\\[-0.3cm]
\end{tabular}\label{tab1}
\end{table*}

\begin{table*}[ht!]\caption[eigenv]{Eigenvalues of the jacobian matrices corresponding to the critical points in table \ref{tab1}.}
\begin{tabular}{@{\hspace{4pt}}c@{\hspace{14pt}}c@{\hspace{14pt}}c@{\hspace{14pt}}
c@{\hspace{14pt}}c@{\hspace{14pt}}c@{\hspace{14pt}}c@{\hspace{14pt}}c}
\hline\hline\\[-0.3cm]
$P_i$ & $\lambda_1$& $\lambda_2$& $\lambda_3$\\[0.1cm]\hline\\[-0.2cm]
%%%%%%%%
$P_1$& $6$ & $2$ & $4$\\[0.2cm]
$P_2$& $2$ & $-2$ & $4$\\[0.2cm]
$P_3$& $6$ & $-\frac{3}{2}+i\frac{\sqrt{15}}{2}$ & $-\frac{3}{2}-i\frac{\sqrt{15}}{2}$\\[0.2cm]
$P_4$& $-2$ & $-4$ & $-4$\\[0.2cm]
$P_5$& $4$ & $6$ & $-4$\\[0.2cm]
$P_6$& $-3$ & $-\frac{3}{2}+\frac{\sqrt{21}}{2}$ & $-\frac{3}{2}-\frac{\sqrt{21}}{2}$\\[0.2cm]
\hline \hline\\[-0.3cm]
\end{tabular}\label{tab2}
\end{table*}

Our aim here is to write the following system of cosmological equations (basically equations (\ref{friedman2}), and (\ref{Raychi2}) with $A=0$):

\bea &&6H^2 f_R=R f_R-f-6H\dot f_R+\rho_m\;,\nonumber\\
&&4\dot H f_R=2H\dot f_R-2\ddot f_R-\rho_m\;,\label{feqs}\eea in the form of an autonomous system of ordinary differential equation (ASODE). To that purpose let us, following Ref.\cite{amendola}, to introduce the dimensionless variables $x$, $y$, $z$:\footnote{Notice that our choice of phase space variables differs from that of Ref. \cite{amendola}, in the definition of one of the variables ($y$).}

\be x=\frac{R}{6H^2}\;,\;\;y=\frac{1}{3\alpha H^2}\;,\;\;z=-\frac{\dot f_R}{H f_R}\;. \label{psv}\ee 

Starting from equations (\ref{friedman2}), (\ref{Raychi2}), (\ref{psv}) and, after a bit of algebra, it is uncomplicated to obtain the following autonomous system -- the right-hand-side of the ODEs do not depend explicitly on the time parameter $\tau$ --, that describes the cosmological dynamics of this model:

\bea &&x'=-2x(x-2)-z(y-2x)\;,\nonumber\\
&&y'=-2y(x-2)\;,\nonumber\\
&&z'=z^2-x z+2(x-2)+\nonumber\\
&&\;\;\;\;\;\;\;\;\;\;\;\;\;\;\;\;\;\;\;\;\;+3(1+x-y+\sqrt{y^2-2xy})\;.\label{asode}\eea Here the comma denotes derivative in respect to the time variable $\tau\equiv\ln a$ (basically the number of e-foldings of inflation). 

The following expressions will be useful: $$q=1-x,\;\;w_{eff}=-\frac{1}{3}(2x-1),$$ where $q=-1-\dot H/H^2$ is the deceleration parameter, and $w_{eff}$ is the effective equation of state parameter for the curvature source in the right-hand-side of Einstein's equations.

The first equation in (\ref{feqs}) can be written in the form of the following (Friedmann) constraint: $$1=\frac{\rho_{m}}{6H^2f_R}+\frac{R}{6H^{2}}-\frac{f}{6f_RH^{2}}-\frac{\dot{f}_R}{Hf_R}\;,$$ or, after the above choice of phase space variables:

\be \bar\Omega_m=1+x-y-z+\sqrt{y^2-2xy}=1-\bar\Omega_{curv}\;,\label{friedmannC}\ee where we have conveniently defined $\bar\Omega_m=\rho_m/6H^2 f_R$, and $\bar\Omega_{curv}$ is the dimensionless curvature (effective or ``dark") energy density.

The next step is to identify the phase space relevant for the cosmological model associated with the autonomous system of ODE (\ref{asode}). Since we are considering positive $\alpha\geq 0\;\Rightarrow\;f_R\geq 0$, then $\bar\Omega_m\geq 0$. Besides, since the expression under the square root in the intermediate expression in (\ref{friedmannC}) must be positive, so that $\bar\Omega_m$ be real, then the following conditions are relevant to define the phase space:

\be 1+x-y-z+\sqrt{y^2-2xy}\geq 0\;,\;\;y\geq 2x\;,\;\;y\geq 0\;,\label{conditions}\ee where the condition $y\geq 2x$ holds only for non-negative $x\geq 0$. For $x<0$ the expression under the square root is always positive. 

Our most precise definition of the corresponding phase space is the following:

\bea &&\Psi_A=\{(x, y, z): \sqrt{y^2-2xy}\geq y+z-x-1\;,\nonumber\\
&&\;\;\;\;\;\;\;\;\;\;\;\;\;\;\;\;\;\;\;\;\;\;\;\;\;\;\;\;\;\;\;\;\;\;\;\;\;\;\;\;\;\;\;\;\;y\geq 0\;,\;\;y\geq 2x\}.\label{PSA}\eea

\begin{figure}[ht]\begin{center}
\includegraphics[width=5.5cm,height=5cm]{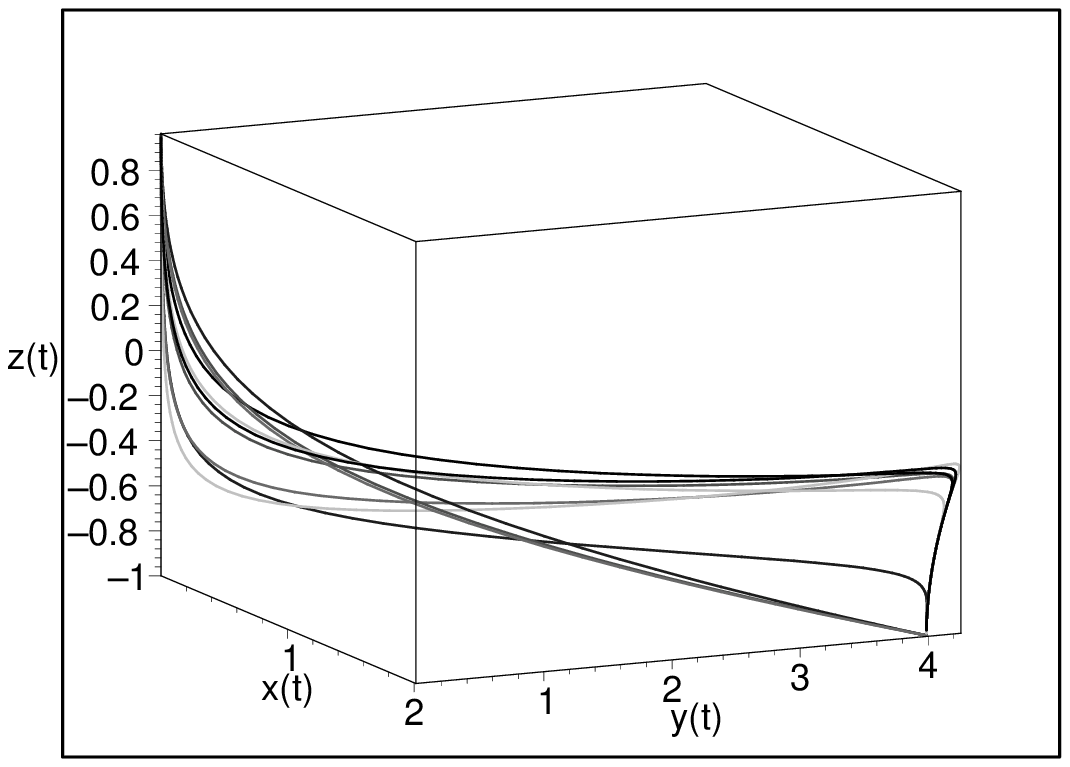}
\includegraphics[width=4cm,height=4cm]{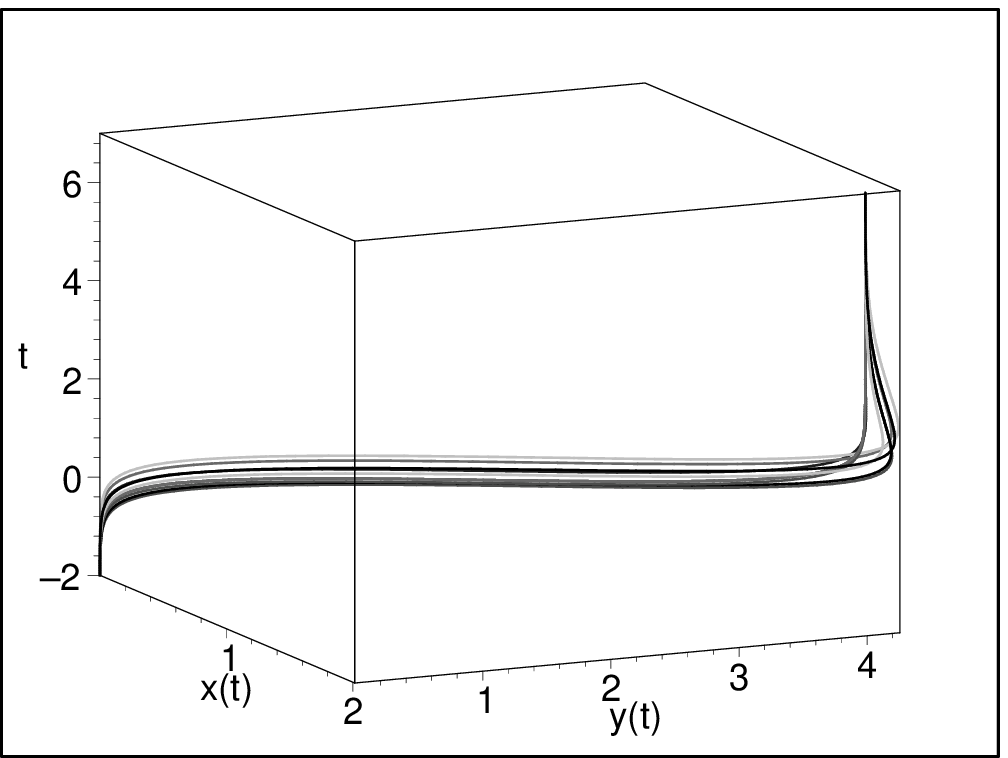}
\includegraphics[width=4cm,height=4cm]{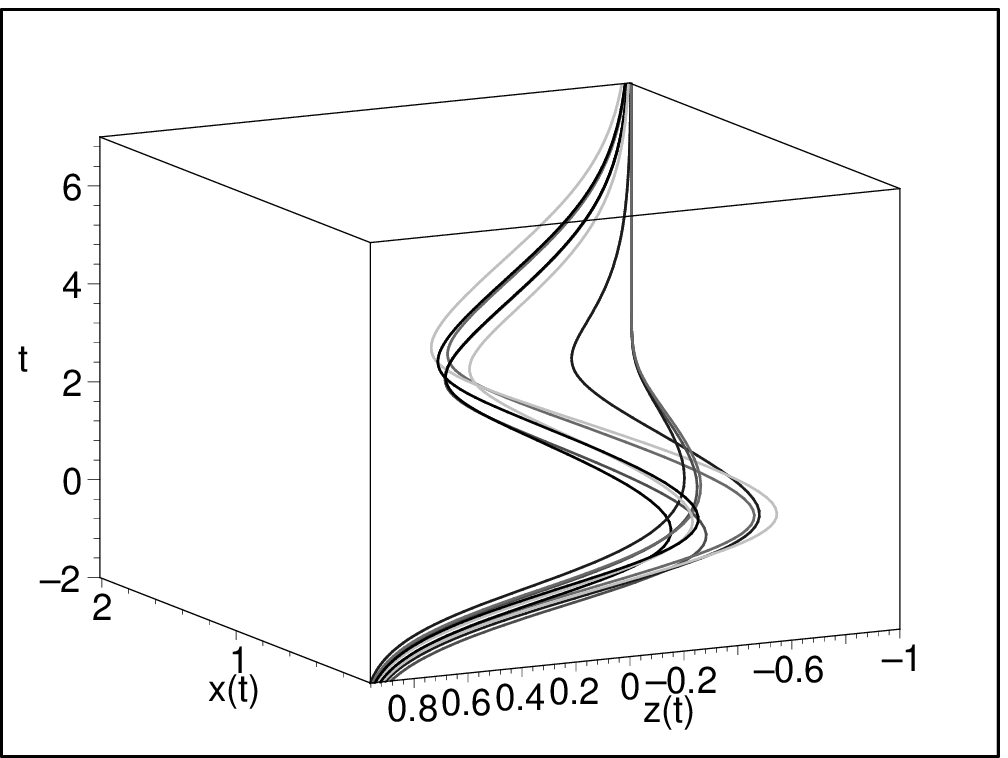}
\vspace{0.3cm}%\bigskip
\caption{Probe paths in phase space originated from different sets of initial conditions for $f(R)=\frac{2}{\kappa^2}\left(1-\sqrt{1-\alpha\kappa^2R}\right)$. The flux in time is shown in the lower panels. The trajectories in $\Psi_A$ emerge from the past attractor (decelerated expansion phase corresponding to the equilibrium point $P_1=(0,0,1)$ in Tab. \ref{tab1}), and are attracted into the future by the curvature dominated inflationary fixed point $P_4=(2,4,-1)$.}\label{fig1}\end{center}\end{figure}

\subsubsection{Equilibrium points}

The next step is to find the roots of the system of equations $x'=0\;,\;\;y'=0\;,\;\;z'=0$. Then one linearizes the ASODE (\ref{asode}) -- means that we expand (\ref{asode}) in the neighborhood of the equilibrium point keeping only the linear terms -- and finds the eigenvalues of the linearization matrix.   

The most relevant properties of the equilibrium points of the autonomous system of ODE (\ref{asode}) are summarized in table \ref{tab1}. Table \ref{tab2} displays the eigenvalues of the respective linearization or Jacobian matrices.

The past attractor in the phase space is the equilibrium point $P_1=(0,0,1)$ in Tab. \ref{tab1}. Although the corresponding cosmological phase is dominated by the curvature ($\bar\Omega_{curv}=1$), the expansion is decelerated ($q=1$). The effective curvature energy density mimics radiation ($w_{eff}=1/3$). This state is characterized by infinitely large values of the Hubble parameter squared $H^2\rightarrow\infty$, and $f_R\propto a(t)$ - the scale factor.

The fixed point $P_2=(0,0,-1)$ in Tab. \ref{tab1} is a saddle equilibrium point. It is associated also with decelerated expansion ($q=1$), but this time there exists a scaling between the energy density of the background fluid $\bar\Omega_m=2$ and that of the effective curvature component $\bar\Omega_{curv}=-1$: $\bar\Omega_m/\bar\Omega_{curv}=-2$. The fact that the dimensionless energy density parameter for the dust $\bar\Omega_m>1$, should not bother us. Recall that we have chosen a convenient definition $\bar\Omega_m=\rho_m/6H^2 f_R$, so that, provided that $2f_R<\Omega_m=\rho_m/3H^2$, we will be faced with $\bar\Omega_m>1$. During this phase of the cosmic evolution the effective energy density of the universe also mimics radiation.

The equilibrium point $P_3=(-1,0,-3)$ is also a saddle and is associated with a scaling of both energy density components: $\bar\Omega_m/\bar\Omega_{curv}=-3/2$. In this case the dynamics of the expansion is ``super"-decelerated ($q=2$), while the effective equation of state parameter mimics that of stiff fluid ($w_{eff}=1$).

The fixed point $P_4=(2,4,-1)$ is the future (late-time) attractor in $\Psi_A$. It is associated with a curvature dominated ($\bar\Omega_{curv}=1$, $\bar\Omega_m=0$), inflationary ($q=-1$) solution. The effective curvature fluid mimics a cosmological constant $\Lambda$ ($w_{eff}=-1$). Another curvature-dominated, inflationary, with $\Lambda$ mimicry, twin-solution is associated with the spiral saddle equilibrium point $P_6=(2,9/2,0)$ in Tab. \ref{tab1}. 

The saddle fixed point $P_5=(2,4,3)$ does not really correspond to a meaningful cosmological scenario. Actually, in this case $\bar\Omega_m=\rho_m/6H^2 f_R=-4$, but since we are considering only non-negative $\alpha\geq 0$, then $f_R=\alpha/\sqrt{1-\alpha R}\geq 0$. Therefore, even if $\bar\Omega_m$ is an effective (convenient) parametrization, it can not be negative.

\subsection{Ghost-free Case ($\mathcal{G}\neq 0$)}

\subsubsection{Autonomous System of ODE}

Now we turn to the more general case when in equations (\ref{friedman2}), (\ref{Raychi2}), the constant $A\neq 0$. In order to transform these equations into an autonomous system of ODE we introduce the following phase space variables:

\bea &&x=\frac{R}{6H^2}\;,\;\;\;\;\;\;y=\frac{1}{3\alpha H^2}\;,\nonumber\\
&&z=-\frac{\dot f_R}{H f_R}\;,\;\;v=-\frac{1}{4AH^2}\;.\label{psv1}\eea Notice that this time the dimension of the phase space has been increased from 3D to a 4-dimensional phase space spanned by the variables $(x,y,z,v)$. However, as we will show quite soon, we can skip one of these variables and, consequently, the phase space corresponding to the present case can be reduced also to a 3-dimensional one.

After the above choice of variables of the phase space, we can write the Friedmann constraint in terms of $x,y,z,v$:

\bea &&\bar\Omega_m=1+x-z-\frac{vy-2}{v-1}+\nonumber\\
&&+\frac{\sqrt{v^2y^2-2v(v-1)xy-2vy}}{v-1}=1-\bar\Omega_{curv}\;,\label{friedmannC1}\eea where now we have conveniently defined $$\bar\Omega_m=\frac{\rho_m}{6H^2 f_R(1+4AH^2)}=\left(\frac{v}{v-1}\right)\frac{\rho_m}{6H^2 f_R}\;.$$ Notice that, contrary to the former case, here the effective matter density parameter $\bar\Omega_m$ can be negative whenever $v<1$. In this case constraints on the domain of $x,y,v$ arise from non-negativity of the expression under the square root: $$y\geq\frac{2(v-1)x+2}{v}\;.$$ 

Other magnitudes of relevance for the analysis that can be straightforwardly put in terms of the above variables are the effective equation of state (EOS) parameter $\omega_{eff}=-1-2\dot{H}/3H^2$, and the deceleration parameter $q=-(1+\dot{H}/H^2)$: $$\omega_{eff}\equiv\frac{1-2x}{3},\;\;q=1-x,$$ respectively.

Following the same procedure as before, after a bit of algebra, we can obtain the following autonomous system of ordinary differential equations out of the cosmological equations (\ref{friedman2}), (\ref{Raychi2}):

\bea &&x'=-2x(x-2)+\frac{2(x-2)^2}{v-1}-\nonumber\\
&&\;\;\;\;\;\;\;\;\;\;\;\;\;\;\;\;\;\;-\frac{z[vy-2(v-1)x-2]}{v-1}\;,\nonumber\\
&&y'=-2y(x-2)\;,\nonumber\\
&&z'=z^2-xz+3z+\frac{2v(x-2)}{v-1}+\nonumber\\
&&\;\;\;\;\;\;\;\;\;\;\;\;\;\;\;\;\;\;\;\;\;\;\;\;+\frac{2z(x-2)}{v-1}+3\bar\Omega_m\;,\nonumber\\
&&v'=-2v(x-2)\;.\label{asode1}\eea 

The 4D phase space relevant for the cosmological model of interest in this case:

\bea &&\Psi_B=\{(x,y,z,v):y\geq 0\;,\;\;v\geq 0\;,\nonumber\\
&&\;\;\;\;\;\;\;\;\;\;\;\;\;\;\;\;\;\;\;\;\;\;\;\;\;\;\;\;\;\;vy\geq 2(v-1)x+2\}\;.\label{PSB}\eea

Worth noticing that in the variables $x,y,z,v$, the particular case ${\cal G}=0$ is recovered from the present case in the formal limit $v\rightarrow\infty$.

\begin{table*}[tbp]\caption[crit]{Properties of the equilibrium points of the autonomous system (\ref{asode1'}). We have defined the parameter $y_*=3/2+k/2-(4\mu-9)/2k$, where $k=(2\mu^2-18\mu+27+2\sqrt{\mu^4-2\mu^3})^{1/3}$.}
\begin{tabular}{@{\hspace{4pt}}c@{\hspace{14pt}}c@{\hspace{14pt}}c@{\hspace{14pt}}c@{\hspace{14pt}}c@{\hspace{14pt}}c@{\hspace{14pt}}c@{\hspace{14pt}}c@{\hspace{14pt}}c}
\hline\hline\\[-0.3cm]
Equilibrium Point &$x$&$y$&$z$& Existence & $\bar\Omega_m$& $\bar\Omega_{curv}$ & $\omega_{eff}$ & $q$\\[0.1cm]
\hline\\[-0.2cm]
%%%%%%%%
$P_1$& $1$&$0$&$0$& $\mu>0$ & $0$&$1$&$-1/3$&$0$\\[0.2cm]
$P_2$& $1$&$0$&$-1$& " &$1$ &$0$ &$-1/3$ &$0$ \\[0.2cm]
$P_3$& $1/2$&$0$&$-3$& " &$5/2$&$-3/2$&$0$&$1/2$ \\[0.2cm]
$P_4$& $3$&$0$&$2$& " &$4$&$-3$&$-5/3$&$-2$ \\[0.2cm]
$P_5^\pm$& $2$&$2\pm\sqrt{4-2\mu}$&$-1$& $0<\mu\leq 2$ &$0$ &$1$&$-1$ &$-1$ \\[0.2cm]
$P_6^\pm$& $2$&$2\pm\sqrt{4-2\mu}$&$3$& $0<\mu\leq 2$ &$-4$ &$5$&$-1$ &$-1$ \\[0.2cm]
$P_7$& $2$&$y_*$ &$0$ & $k>0\;(\mu>2.25)$ &$0$ &$1$ &$-1$ &$-1$ \\[0.2cm]
\hline \hline
\end{tabular}\label{tab3}
\end{table*}
\begin{table*}[tbp]\caption[eigenv]{Eigenvalues of the linearization matices corresponding to the critical points in table \ref{tab3}. For the equilibrium point $P_7$ the analytic expressions of the eigenvalues are extremely huge and complex, so that we have decided to show numeric results for several values of the control parameter $\mu$ instead.} \begin{tabular}{@{\hspace{4pt}}c@{\hspace{14pt}}c@{\hspace{14pt}}c@{\hspace{14pt}}c@{\hspace{14pt}}c}
\hline\hline\\[-0.3cm]
Equilibrium Point & $\mu$ & $\lambda_1$& $\lambda_2$& $\lambda_3$\\[0.1cm]\hline\\[-0.2cm]
%%%%%%%%
$P_1$& $\mu>0$ & $1$&$2$&$4$\\[0.2cm]
$P_2$& " & $-1$&$2$&$2$\\[0.2cm]
$P_3$& " & $3$&$-3/4+i\sqrt{71}/4$&$-3/4-i\sqrt{71}/4$\\[0.2cm]
$P_4$& " & $-2$&$-4$&$-5$\\[0.2cm]
$P_5^\pm$& $0<\mu\leq 2$ & $-2$&$-4$&$-4$\\[0.2cm]
$P_6^\pm$& $0<\mu\leq 2$ & $4$&$6$&$-4$\\[0.2cm]
$P_7$ & $5$ & $2.1$ & $-4.03+0.78 i$ & $-4.03-0.78 i$ \\[0.2cm]
     & $10$ & $3.26$ & $-4.63+0.8 i$ & $-4.63-0.8 i$ \\[0.2cm]
     & $15$ & $4.19$ & $-5.1+0.55 i$ & $-5.1+0.55 i$ \\[0.2cm]
\hline \hline
\end{tabular}\label{tab4}
\end{table*}

\subsubsection{Parametric ASODE}

It is noticeable the similarity between 2nd and 4th equations in (\ref{asode1}). However, by looking at the definitions of the variables $y$ and $v$
one sees that this is not casual. Actually, from (\ref{psv1}) one can see that 

\be y=\mu\;v\;,\;\;\mu\equiv\frac{4\beta}{3\alpha^2}.\label{parameter}\ee Hence, the ASODE (\ref{asode1}) is not really a 4-dimensional system of equations but a 3-dimensional one. In fact it is a 3D, one-parameter ASODE. In the theory of the dynamical systems the parameter $\mu\equiv 4\beta/3\alpha^2$ is called as ``control parameter". It usually controls the occurrence of bifurcations -- a change of the value of the control parameter produces a change in the topology of the phase portrait.\footnote{Boundaries between regions of phase portrait with different topology are called as ``bifurcation sets".} That such bifurcations arise in the present study will be clear when we found that the asymptotic structure of the DBI-EH model (${\cal G}=0\;\Rightarrow\;\mu=0$) and that of the general, ghost-free case ($\mu\neq 0$, $\mu>0$), are qualitatively different.

In what follows we will skip the variable $v$, and will write the 4D autonomous system (\ref{asode1}) as a 3D, one-parameter ASODE in the variables $x,y,z$:

\bea &&x'=-2x(x-2)+\frac{2\mu (x-2)^2}{y-\mu}-\nonumber\\
&&\;\;\;\;\;\;\;\;\;\;\;\;\;\;\;\;\;\;\;\;\;\;\;\;\;\;-\frac{z[y^2-2xy+2\mu(x-1)]}{y-\mu}\;,\nonumber\\
&&y'=-2y(x-2)\;,\nonumber\\
&&z'=z^2-xz+3z+\frac{2y(x-2)}{y-\mu}+\nonumber\\
&&\;\;\;\;\;\;\;\;\;\;\;\;\;\;\;\;\;\;\;\;\;+\frac{2\mu z(x-2)}{y-\mu}+3\bar\Omega^\mu_m(x,y,z),\label{asode1'}\eea where

\bea &&\bar\Omega^\mu_m(x,y,z)=1+x-z-\frac{y^2-2\mu}{y-\mu}+\nonumber\\
&&\;\;\;\;\;\;\;\;\;\;\;\;\;\;\;\;\;\;\;\;\;\;\;\;+\frac{y\sqrt{y^2-2xy+2\mu(x-1)}}{y-\mu}\;.\label{friemannC1'}\eea The one-parameter, 3D phase space spanned by the variables $x,y,z$ -- relevant for the study of the asymptotic properties of the cosmological model (\ref{friedman2},\ref{Raychi2}) -- can be finally defined in the following way:

\be \Psi^\mu_B=\{(x,y,z):y\geq 0\;,\;y^2-2xy+2\mu(x-1)\geq 0\}\;.\label{PSBm}\ee

We want to emphasize that the bound 

\be {\cal B}(x,y,\mu)\equiv y^2-2xy+2\mu(x-1)\geq 0\;,\label{bound}\ee has to be necessarily satisfied in order for the density parameter to be a real quantity.

\begin{figure}[t!]\begin{center}
\includegraphics[width=8cm,height=6.5cm]{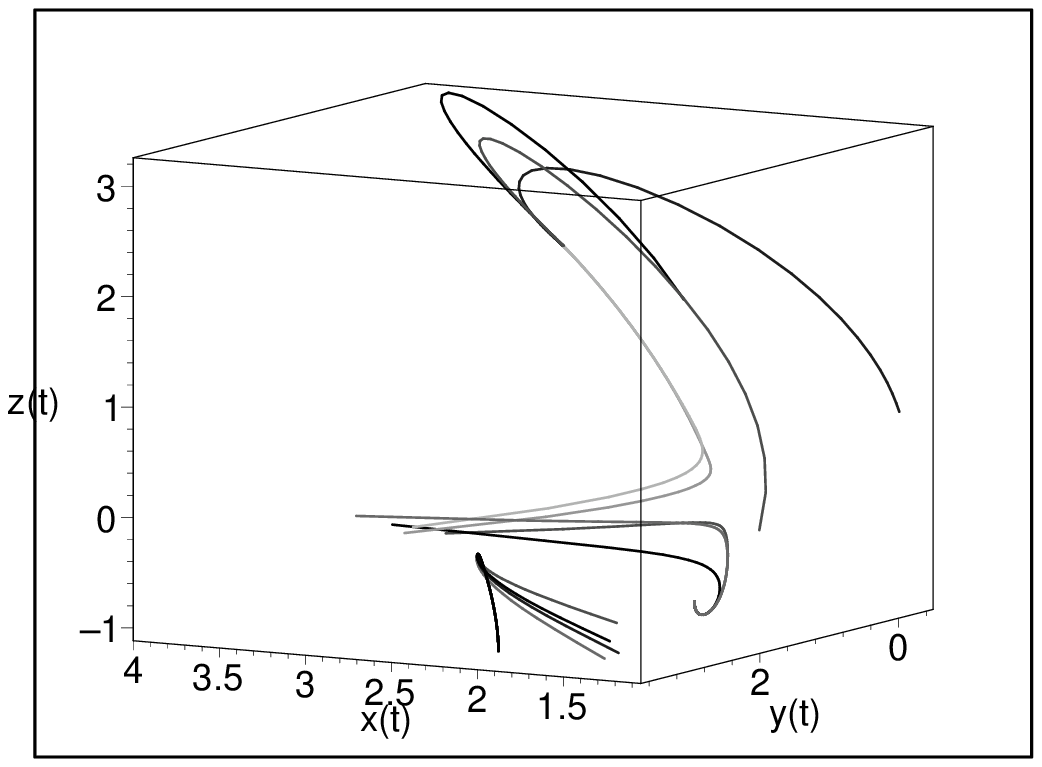}
\includegraphics[width=4cm,height=3.5cm]{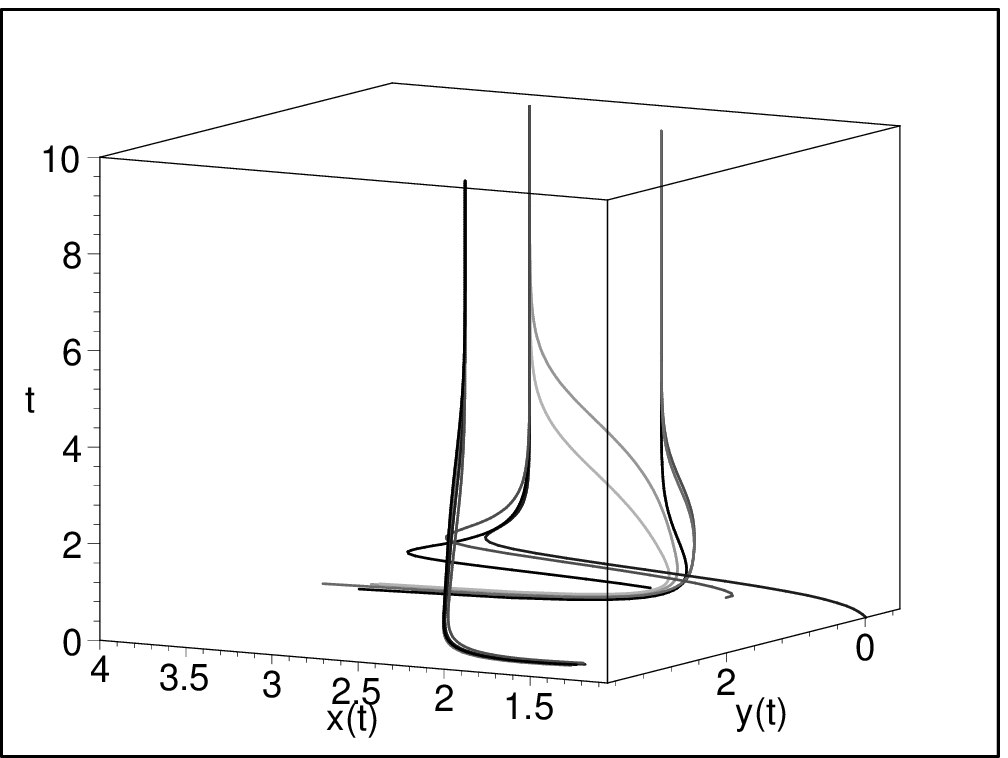}
\includegraphics[width=4cm,height=3.5cm]{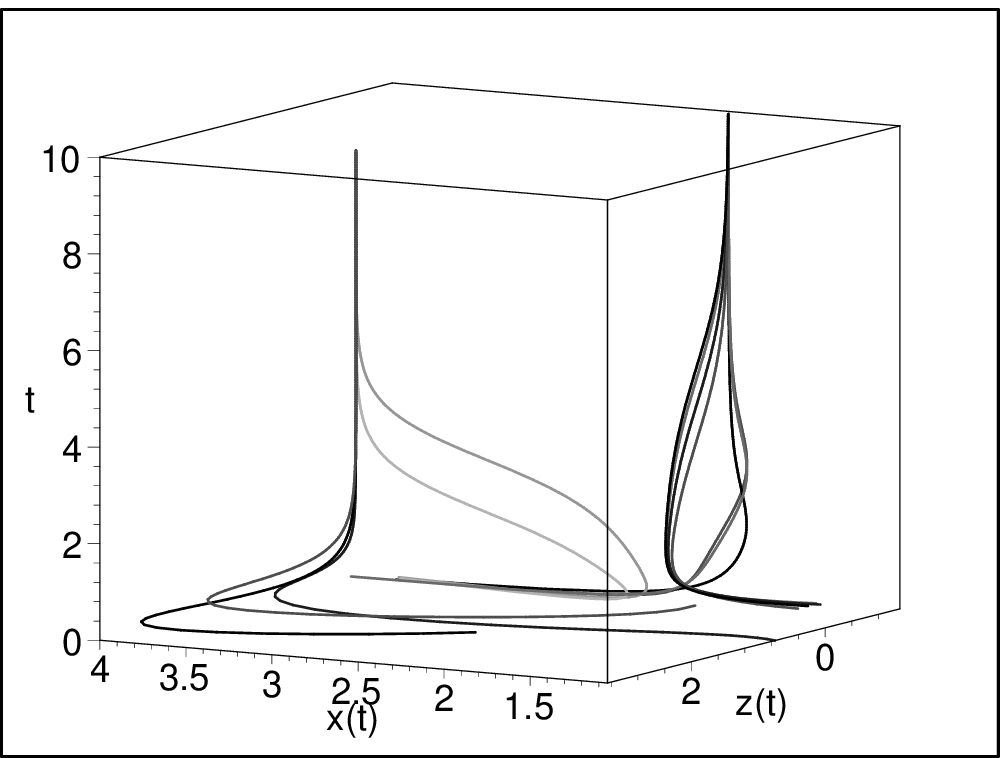}
\vspace{0.3cm}%\bigskip
\caption{Drawing of the phase space to show the ``multi-attractor" structure ($\mu=1$). This future asymptotic structure is better seen in the $\tau$-flow drawings in the lower panels.}\label{fig2}\end{center}\end{figure}

\subsubsection{Equilibrium points}

There are found eight equilibrium points corresponding to the autonomous system of ODE (\ref{asode1'}) in the phase space $\Psi^\mu_B$. Their most relevant properties are listed in the table \ref{tab3}, while the eigenvalues of the linearization matrices corresponding to these points -- invaluable as they are to judge about their stability --, are displayed in table \ref{tab4}. 

Notice that the point $P_3$ in Tab. \ref{tab3} is not really a fixed point of (\ref{asode1'}), since the bound (\ref{bound}) is not satisfied. Actually, for $P_3=(1/2,0,-3)$; $${\cal B}(x,y,\mu)=-\mu<0\;,$$ since we are considering positive $\mu$-s.

Fixed points $P_1$, $P_2$, and $P_4$ are associated with infinitely large values of the Hubble parameter $H\rightarrow\infty$, while points $P_5$-$P_7$ correspond to de Sitter cosmological phases with (different) finite Hubble rates. 

The main asymptotic properties of the model with $f(R)=2(1-\sqrt{1-\alpha R+\beta {\cal G}})$ can be summarized as it follows:

\begin{itemize}

\item The equilibrium point $P_1$ is associated with curvature-dominated ($\bar\Omega_{curv}=1$, $\bar\Omega_m=0$), non-inflationary dynamics as long as $q=0$. It is the past attractor in $\Psi^\mu_B$. 

\item Another non-inflationary phase can be associated with the fixed point $P_2$, corresponding to matter domination ($\bar\Omega_m=1$, $\bar\Omega_{curv}=0$). It is a saddle equilibrium point in $\Psi^\mu_B$.

\item The scaling solution $\bar\Omega_m/\bar\Omega_{curv}=-5/3$ -- point $P_3$ -- is associated with decelerated expansion ($q=1/2$). The effective ``curvature" fluid mimics dust ($w_{eff}=0$). However, as already noticed this is not a fixed point of (\ref{asode1'}) since it does not satisfy the bound (\ref{bound}).

\item Super-inflationary dynamics can be associated with the equilibrium point $P_4$ in Tab \ref{tab3} (also a scaling solution since $\bar\Omega_m/\bar\Omega_{curv}=-4/3$). Actually, in this case $w_{eff}=-5/3<-1$ ($q=-2$). Otherwise, the effective fluid mimics phantom behavior, which can be explained here as a curvature effect (no phantom fields at all).  Point $P_4$ is a future attractor in $\Psi^\mu_B$ (this time not a global one, see below). It corresponds to a singular state (infinite curvature $R\rightarrow\infty$) that could be identified with a catastrophic fate inherent in a large class of phantom models.

\item Curvature-dominated ($\bar\Omega_{curv}=1$, $\bar\Omega_m=0$) equilibrium points $P_5^\pm$ -- corresponding to de Sitter inflationary phases of the cosmic expansion since $q=-1$ -- are also future attractors in $\Psi^\mu_B$. These points exist whenever $\mu\leq 2$.

\item Points $P_6^\pm$ -- associated also with de Sitter (inflationary) expansion -- are saddle fixed points in the phase space (\ref{PSBm}). These can be linked with scaling solutions since $\bar\Omega_m/\bar\Omega_{curv}=-4/5$. The fact that the effective dimensionless matter energy density parameter $\bar\Omega_m$ is a negative parameter, is not a problem and is easily explained as due to our definition: $$\bar\Omega_m=\left(\frac{v}{v-1}\right)\frac{\rho_m}{6H^2 f_R}=\left(\frac{y}{y-\mu}\right)\frac{\rho_m}{6H^2 f_R}.$$ Actually, for $y<\mu$ ($v<1$) $\bar\Omega_m<0$. Points $P_6^\pm$ exist only if $\mu\leq 2$.

\item It is found, additionally, another de Sitter saddle equilibrium configuration (point $P_7$) in $\Psi^\mu_B$. The fixed point $P_7$ can be associated also with a curvature-dominated solution ($\bar\Omega_{curv}=1$, $\bar\Omega_m=0$) of the cosmological equations (\ref{friedman2}), (\ref{Raychi2}). This equilibrium point exists whenever the following inequality holds: $$(2\mu^2-18\mu+27+2\sqrt{\mu^4-2\mu^3})^{1/3}>0.$$ Numerical investigations show that this bound is equivalent to the following restriction on the values the control parameter $\mu$ can take: $\mu>2.25$.

\end{itemize}

We want to emphasize several interesting features of the asymptotic structure of the model (\ref{friedman2},\ref{Raychi2}). Perhaps the most interesting fact is the ``multi-attractor" structure into the future of the $\tau$-development of (\ref{asode1'}) for values of the control parameter $\mu\leq 2$. In this case there co-exist three future attractors (equilibrium points $P_4$, $P_5^+$, and $P_5^-$). Otherwise, for this interval of control parameter space there is no a unique global late-time attractor. Anyway, judging by the magnitudes of the eigenvalues, convergence of phase space trajectories towards the point $P_4$ is stronger than towards points $P_5^\pm$. 

Not of less importance is the existence of a saddle equilibrium point correlated with matter dominance -- point $P_2$ ($\bar\Omega_m=1$, $\bar\Omega_{curv}=0$). The existence of this point is crucial to explain the formation of structure. Existence of an inflationary de Sitter saddle equilibrium point $P_7$ could be important to explain early time inflation.

Last but not least: the super-inflationary attractor point $P_4$ (a global future attractor if $\mu>2$) represents another example of how to mimic catastrophic phantom behavior ($w_{eff}<-1$) without any phantom fields. It is evident that in the model crossing of the phantom divide is possible (it is in fact unavoidable for $\mu>2$).

\begin{figure}[t!]\begin{center}
\includegraphics[width=4cm,height=3.5cm]{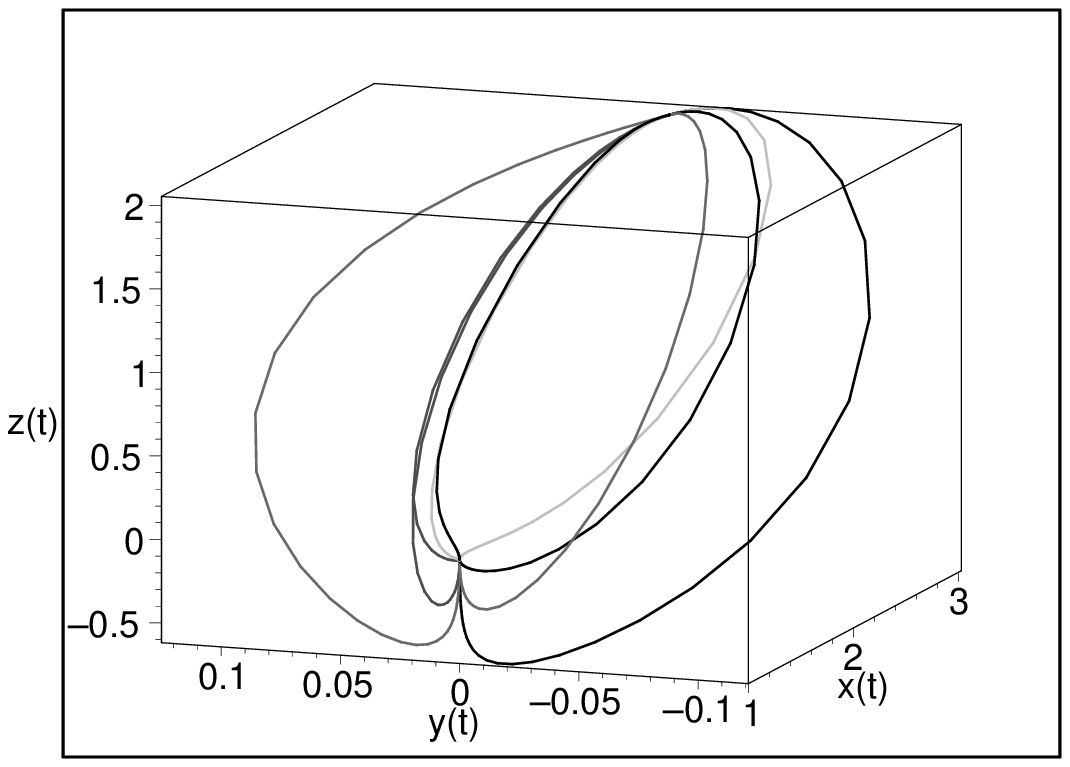}
\includegraphics[width=4cm,height=3.5cm]{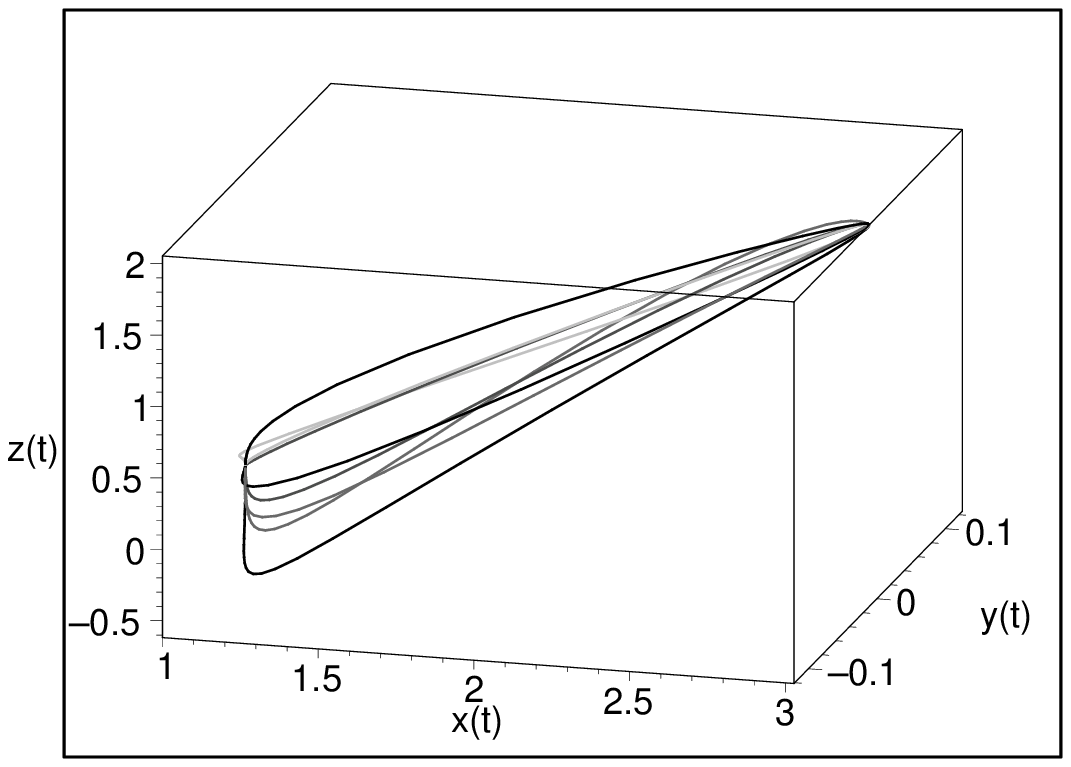}
\includegraphics[width=4cm,height=3.5cm]{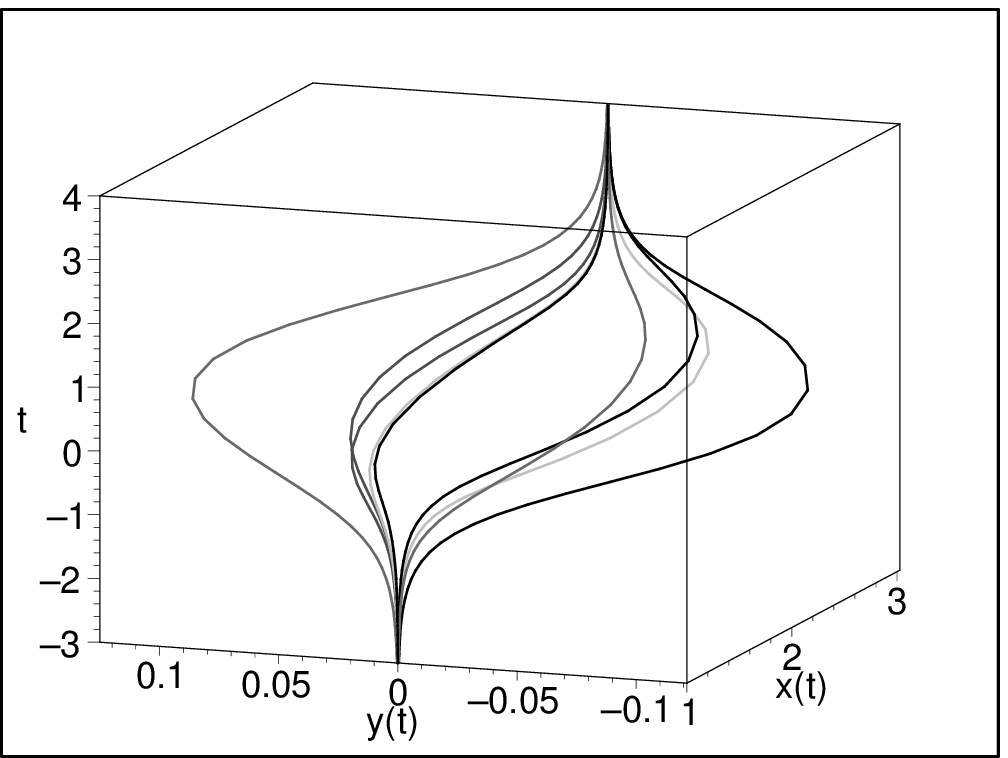}
\includegraphics[width=4cm,height=3.5cm]{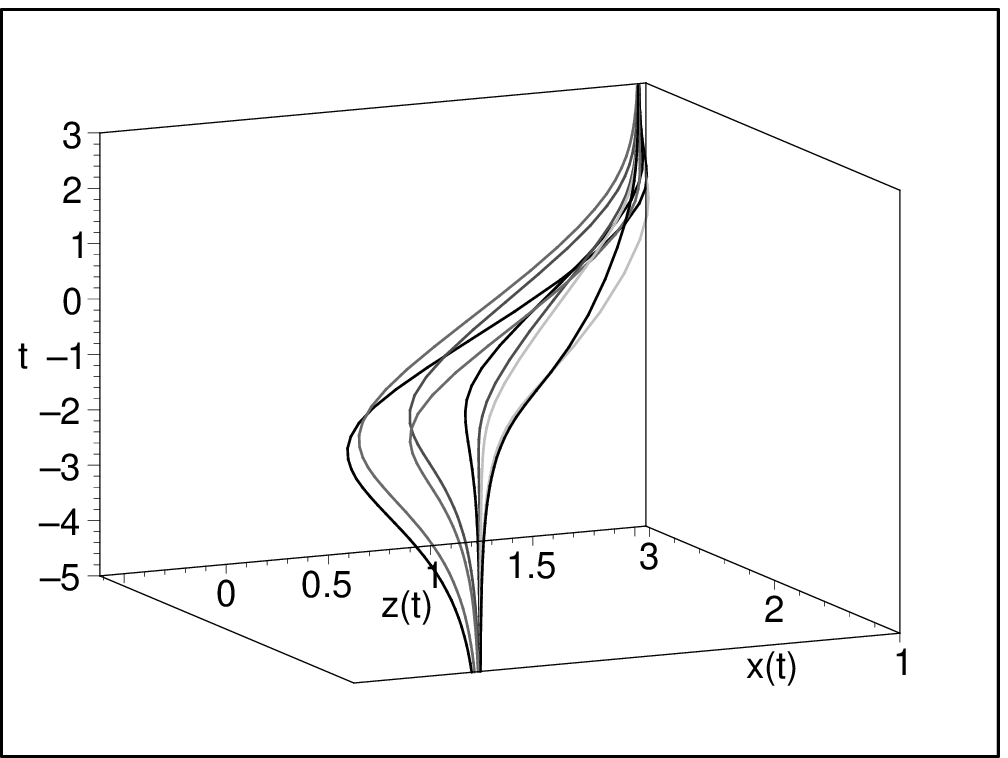}
\vspace{0.3cm}%\bigskip
\caption{Probe paths in phase space originated from different sets of initial conditions for $f(R)=\frac{2}{\kappa^2}\left(1-\sqrt{1-\alpha\kappa^2R+\beta{\cal G}}\right)$. The control parameter $\mu=4\beta/3\alpha^2$ has been chosen to be $\mu=5$ so that points $P_5^\pm$ and $P_6^\pm$ do not exist. The flux in time is shown in the lower panels. The trajectories in $\Psi^\mu_B$ converge towards the global (``super-inflationary") future attractor -- equilibrium point $P_4=(3,0,2)$ in Tab. \ref{tab3}. The past attractor is the curvature-dominated (non-inflationary) solution -- point $P_1=(1,0,0)$ in Tab. \ref{tab3}.}\label{fig3}\end{center}\end{figure}

\begin{figure}[t!]\begin{center}
\includegraphics[width=4cm,height=3.5cm]{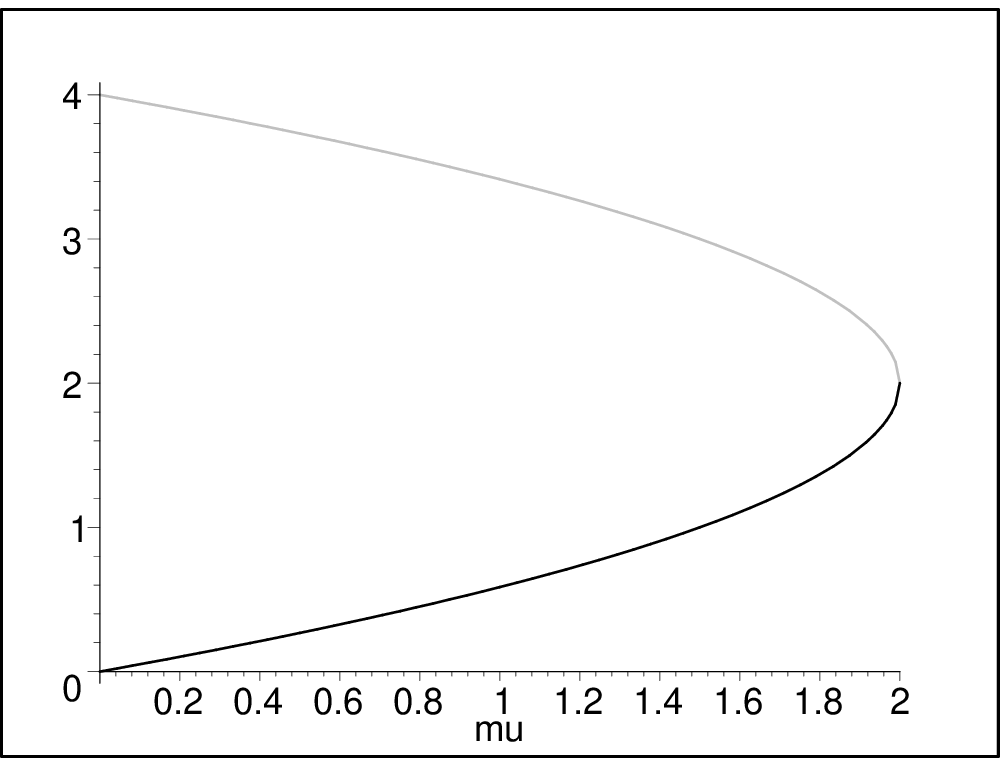}
\includegraphics[width=4cm,height=3.5cm]{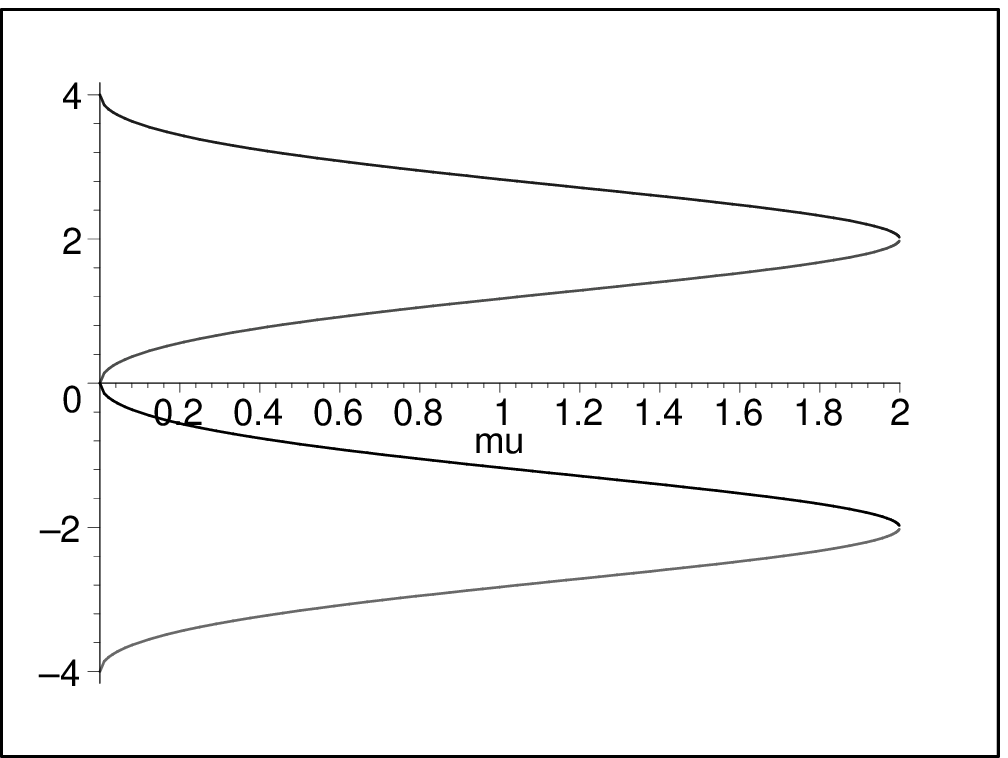}
\vspace{0.3cm}%\bigskip
\caption{Plots of the phase space coordinate $y$ vs $\mu$ for the equilibrium points $P_5^\pm$ and $P_6^\pm$ ($\mu\leq 2$) -- left panel. In the right panel it is shown a plot of the effective dimensionless matter density parameter $\bar\Omega_m$ vs $\mu$ for points $P_5^\pm$ (positive half), and $P_6^\pm$ (negative half). At $\mu=2$ points $P_5^+$ and $P_5^-$ (also $P_6^+$ and $P_6^-$) coincide, so that there are 7 equilibrium points in $\Psi_B^\mu$ instead of 9. For $\mu>2$ these points do not exist at all (only 5 equilibrium points are found). This is a clear illustration of the bifurcation in the control parameter.}\label{fig4}\end{center}\end{figure}

\subsubsection{Bifurcations}

To get additional information on the asymptotic dynamical properties of the model under study one can rely on drawings of the phase space, where probe paths evolved from given (arbitrary) initial data probe the phase space and, eventually, uncover the nature of the relevant equilibrium points. Here, as in the former subsection, we make a combined use of the standard tools of the (linear) dynamical systems analysis and of phase space drawings. This time the phase space drawings will serve as a graphic illustration of what it has been said in the former subsection (what is shown in tables Tab. \ref{tab3} and Tab. \ref{tab4}). They clearly illustrate, in particular, bifurcations in respect to the control parameter $\mu$.

In the upper part of figure \ref{fig2} probe trajectories originated from different initial data, with control parameter $\mu=1$, converge (increasing $\tau$) into three different attractor points: $P_5^+=(2,3.41,-1)$ (bunch of trajectories closer to the reader), $P_5^-=(2,0.6,-1)$ (bunch to the center of the drawing), and $P_4=(3,0,2)$ (set of trajectories at the end). This is more evident in the $\tau$-flow diagrams in the lower part of the figure. In the left panel the convergence towards three different points is apparent. In the right panel only two points seem to be points of convergence of the phase trajectories. Note, however, that in this diagram only the coordinate axes $x$ and $z$ are displayed. Hence, since the points $P_5^+$ and $P_5^-$ differ only in the coordinate $y$, in this drawing they appear as a single convergence point (the one closer to the reader).

In figure \ref{fig3} a different value of the control parameter $\mu=5$ has been chosen. In this case, since $\mu>2$, equilibrium points $P_5^\pm$ (also $P_6^\pm$) do not occur. Consequently there is only a (global) future attractor: the point $P_4$. As seen from Fig. \ref{fig3} (upper part) probe paths in $\Psi_B^\mu$ converge into the past (decreasing $\tau$) towards the past attractor (equilibrium point $P_1$), while these converge into the future towards the attractor point $P_4$. In the right panel a view from a different angle is shown. This asymptotic structure is confirmed by the $\tau$-flow diagrams in the lower part of the figure.

The above figures (Fig. \ref{fig2} and Fig. \ref{fig3}) illustrate the bifurcation in respect to the control parameter $\mu$. Depending on the magnitude of the parameter $\mu$ there are 9 ($\mu<2$), 7 ($\mu=2$), or 5 ($\mu>2$) equilibrium points in the phase space $\Psi_B^\mu$. Therefore, $\mu=2$ is the critical value or the bifurcation point. 

In the figure Fig. \ref{fig4} we show bifurcation diagrams $y$ vs $\mu$ -- left panel, and $\bar\Omega_m$ vs $\mu$ -- right panel, at the equilibrium points $P_5^\pm$ and $P_6^\pm$. It is apparent that as $\mu\rightarrow 2\;\Rightarrow\;y\rightarrow 2$, points $P_5^+$ and $P_5^-$ tend to be a single equilibrium point ($P_5=(2,2,-1)$). The same is true for the points $P_6^+$ and $P_6^-$: at the critical value $\mu=2$ these are a single fixed point $P_6=(2,2,3)$. In this case the four equilibrium points $P_5^\pm$ and $P_6^\pm$ reduce to just two equilibrium points $P_5$ and $P_6$. For values $\mu>2$ (not drawn in the figure) even $P_5$ and $P_6$ are erased from the phase space. 

In the present model bifurcations in the control parameter space demonstrate the strong dependence of the dynamic properties of the model on the values of the overall parameters $\alpha$ and $\beta$.

\section{Discussion}

Knowledge of the equilibrium points in the phase space corresponding to a given cosmological model is a very important information since, independent on the initial conditions chosen, the orbits of the corresponding autonomous system of ODE will always evolve for some time in the neighborhood of these points. Besides, if the point were a stable attractor, independent on the initial conditions, the orbits will always be attracted towards it (either into the past or into the future). Going back to the original cosmological model, the existence of the equilibrium points can be correlated with generic cosmological solutions that might really decide the fate and/or the origin of the cosmic evolution. In a sense the knowledge of the asymptotic properties of a given cosmological model is more relevant than the knowledge of a particular analytic solution of the corresponding cosmological equations. While in the later case one might evolve the model from given initial data giving a concrete picture that can be tested against existing observational data, the knowledge of the asymptotic properties of the model gives the power to realize which will be the generic behavior of the model without solving the cosmological equations. In the dynamical systems language, for instance, a given particular solution of the Einstein's equations is just a single point in the phase space. Hence, phase space orbits show the way the model drives the cosmological evolution from one particular solution into another one. Equilibrium points in the phase space will correspond to solutions of the cosmological (Einstein's) equations that, in a sense, are preferred by the model, i. e., are generic. The lack of equilibrium points that could be correlated with a given analytic solution of the model, amounts to say that this solution is not generic, otherwise it can be attained under a very carefully arrangement of the initial conditions only.

The asymptotic structure of the model (\ref{friedman2},\ref{Raychi2}) (or (\ref{friedman},\ref{Raychi})) is extremely rich and complex. Contrary to what is believed (see, for instance, Ref. \cite{amendola}), there are found in this kind of $f(R,{\cal G})$-theory, equilibrium points that can be associated with matter-dominated and matter-curvature scaling solutions. This is a very nice feature that makes the model attractive to do cosmology, since, by appropriately arranging the free parameters, there is room to accommodate the amount of structure observations confirm it exists. The fact that, for $\mu>2.25$ (see existence conditions in Tab. \ref{tab3}), two inflationary solutions with different Hubble rates, and with different stability properties, co-exist together: i) the inflationary de Sitter phase linked with the saddle fixed point $P_7$; $$H^2=\frac{2k}{3\alpha(k^2+3k-4\mu+9)}\;,$$ where $k\equiv (2\mu^2-18\mu+27+2\sqrt{\mu^4-2\mu^3})^{1/3}$, and ii) the super-inflationary ghost-like phase ($w_{eff}<-1$, $q=-2$) -- point $P_4$ -- with infinite Hubble rate, is encouraging as well since, one can try to accommodate both early and late time inflationary stages into a united picture, where inflation is driven by curvature effects. In fact, the latter solution can be associated with late-time speedup of the cosmic expansion, while the former one might be linked with early inflation as long as, due to its stability properties, it is a transient phase. By appropriately choosing the value of the free parameter $\mu\equiv 3\alpha^2/4\beta$, one can manage to produce the necessary amount of inflation at early times, besides, since $P_7$ is a saddle critical point, then, exit from this inflationary stage is natural. Actually, fueled by the stability properties of the corresponding cosmological phase, the dynamics forces the model to leave the saddle critical point $P_7$ to, eventually, approach to the global late-time super-inflationary attractor $P_4$ (recall that we are considering $\mu>2.25$, so that the de Sitter attractors $P_5^\pm$ do not exist).\footnote{Similar results about the possibility to have both inflationary quintessential and super-inflationary ghost-like phases in a model have been obtained for more elaborated models \cite{saridakis}.} For $\mu\leq 2$, the possibility to explain early and late time inflation in a united scheme is based on the co-existence of the super-inflationary attractor $P_4$ and of the inflationary (de Sitter) saddle critical points $P_6^\pm$.\footnote{Notice that, even if there co-exist de Sitter attractors -- points $P_5^\pm$, and de Sitter solutions associated with saddle points $P_6^\pm$, there is no room for a united description of early and late time inflationary stages of the cosmic expansion in this case, since the Hubble rate is the same for both asymptotic stages.}

The existence of the non-inflationary, curvature dominated equilibrium point $P_1=(1,0,0)$, being always the past attractor in the phase space, allows us to associate the starting point of every probe path in the phase space with a state characterized by infinitely large values of the Hubble parameter, a state usually linked with the big bang singularity. Therefore, contrary to what one should expect, the initial singularity is a generic feature of the model sourced by the action (\ref{dbi action}).\footnote{This kind of statements has to be taken with caution since the dynamical systems analysis uncovers the asymptotic properties of the dynamics. Hence, for instance, the asymptotic state where $H^2$ blows up might be reach in an infinite time, in which case one would not want to say that the space-time is properly singular.} In case such a singularity were removed by the non-linear DBI dynamics, as expected, the starting point of the cosmic dynamics were to be linked with a regular solution resembling a finite big-bang event, which does not seem to be the case. It has to be pointed out, however, that, even if the generic behavior is associated with a big-bang singularity, there might exist particular solutions that are regular at the beginning of the cosmic expansion. To show that the latter statement is indeed true, let us write the combination $\alpha(R+A{\cal G})$ in the action (\ref{dbi f}, \ref{F}), in terms of the variables of the phase space: $$\alpha(R+A{\cal G})=\frac{2(y-\mu)x+2\mu}{y^2}\;.$$ There is a class of trajectories in the $x-y$ phase plane: $$x=\frac{\mu}{\mu-y}+\frac{y^2 f(y)}{\mu-y}\;,$$ where $f(y)$ is a generic function such that, in the limit $y\rightarrow 0$, it is finite (note from the above expression that, at $y=0$, $x=1$). For these particular trajectories as one approaches the point $P_1=(1,0,0)$, the above curvature invariant $$\alpha(R+A{\cal G})=-f(y)\;,$$ so that, at $x=1$, $y=0$ it is finite, otherwise, the action is finite and there is no singularity associated with the past attractor $P_1$.

\section{Conclusion}

In the present paper we have performed a throughout study of the asymptotic properties of a cosmological model that is based on a non-linear (DBI-type) modification of Einstein-Hilbert general relativity previously proposed in \cite{comelli}. The modification includes, in addition to the standard Einstein-Hilbert term within the square root in the action, also the Gauss-Bonnet invariant. Theories of this kind are oriented to remove the pathological spacetime singularities from the original EH theory. In former works in the bibliography the present model has been demonstrated to be ghost-free and, for a given interval of the space of parameters, it is also free of the Coulomb-like Schwarszchild singularity \cite{comelli}. 

The study has been performed in two steps: i) first we explored the asymptotic properties of a simpler model where the Gauss-Bonnet invariant is absent, and then ii) we studied properly the model of interest. The idea was to solve first a related case easier to handle so that we could get some insight into the more complicated situation. Even if the DBI model without the GB invariant has a ghost (spin-zero) degree of freedom \cite{qln}, so that it is not field-theoretically consistent (negative-norm states arise when invoking quantum computations), the study showed, nonetheless, that even this simpler -- theoretically inconsistent -- model has very nice asymptotic features that make it attractive to do cosmology.

In general the asymptotic structure of the Friedmann-Robertson-Walker cosmological model originated from the action (\ref{dbi action}) is extremely complex and rich. There can be found equilibrium points associated with matter-dominated, matter-curvature scaling, de Sitter and even phantom-like solutions. The multi-attractor structure in the future $\tau$-development of the autonomous system of ODE (\ref{asode1'}) is a very interesting finding. Even more interesting are the bifurcations in the control parameter $\mu$-space, showing the strong dependence of the model upon the values of the overall parameters $\alpha$, and $\beta$  (this is very convenient as long as makes the model very flexible to accommodate different kinds of cosmological dynamics). 

Our results concerning the occurrence of a future singularity as well as of a past big-bang singularity, confirm the results of previous studies on this subject within a large class of modified theories of gravity \cite{odi1}. Otherwise stated, contrary to what one should expect from a non-linear modification of general relativity such as the one based on the action (\ref{dbi action}), removal of the big bang singularity is not a generic feature of the corresponding cosmological model. Instead, the starting point of the cosmic expansion is a state with infinitely large values of the Hubble parameter squared $H^2\rightarrow\infty$, which is associated with a past attractor in the phase space. Additionally, there is a future super-inflationary attractor associated with a finite-time singularity of the kind inherent in a large class of phantom models of dark energy.

We thak S D Odintsov, E Saridakis, and F Fiorini for useful comments and for calling our attention upon references \cite{odi}, \cite{odi1}, and \cite{saridakis}. This work was partly supported by CONACyT M\'{e}xico, under grants 49865-F, 54576-F, 56159-F, 49924-J, 105079, 52327, and by grant number I0101/131/07 C-234/07, Instituto Avanzado de Cosmologia (IAC) collaboration. R G-S acknowledges partial support from COFAA-IPN and EDI-IPN grants, and SIP-IPN 20090440. T G, Y N and Y L acknowledge the MES of Cuba for partial support of the research.

\end{document}